\documentstyle[preprint,aps,epsf,amsmath,amssymb,graphicx]{revtex}
\begin{document}
\preprint{THEF-NYM 02.01/KVI-1564}
\draft
\tighten

\title{Angle-dependent normalization of neutron-proton \\
       differential cross sections}

\author{J.J. de Swart$^a$ and R.G.E. Timmermans$^b$}

\address{$^a$Institute for Theoretical Physics, University of Nijmegen,
         P.O. Box 9010, \\ 6500 GL Nijmegen, The Netherlands}
\address{$^b$Theory Group, KVI, University of Groningen,
         Zernikelaan 25, \\ 9747 AA Groningen, The Netherlands}

\date{\today}

\maketitle

\begin{abstract}
Systematic errors in the database of $np$ differential cross sections
below 350 MeV are studied. 
By applying angle-dependent normalizations 
with the help of the energy-dependent Nijmegen partial-wave analysis PWA93,
the $\chi^2$-values of some 
seriously flawed data sets can be reduced significantly
at the expense of a few degrees of freedom.
It turns out that in these special cases
the renormalized data sets can be made statistically acceptable such that 
they do not have to be discarded any longer in partial-wave analyses
of the two-nucleon scattering data.
\end{abstract}


\section{Introduction}
A measurement of the differential cross section for elastic
neutron-proton ($np$) scattering is notoriously difficult. 
It is even so difficult that almost none of the data sets measured at
energies below the pion-production threshold is completely free of
systematic flaws. In partial-wave analyses (PWA's) of the $np$
scattering data~\cite{Sto93,Arn00} some of these flaws do not give
rise to sizable systematic contributions to $\chi^2$. Data sets with
such minor flaws will not distort too much the statistics~\cite{Ber88}
in these PWA's, and therefore such sets can be included in the database; 
examples are the LAMPF data~\cite{Bon78} and the TRIUMF data~\cite{Kee82}.
Some flaws, on the other hand, are so serious that their contribution
to $\chi^2$ dominates over the statistical contribution to the extent 
that the standard rules of statistics no longer apply.
Consequently, such data sets~\cite{She74,Hue80,Eri95,Rah98,Fra00}
must be excluded from the databases used in PWA's. This is of course
an unfortunate and undesirable situation, especially in view of the
waste of investment and effort involved in these experiments.

In this paper we present the ``adnorm'' method. 
This is a method of Angle-Dependent NORMalization~\cite{Adnor} 
to treat certain systematically-flawed $np$ differential cross sections.
This adnorm method is meant to be used only in good, energy-dependent PWA's.
We will show that the application of this method to certain data sets, 
which were previously unacceptable in energy-dependent PWA's, 
can give impressive results. 
The values of $\chi^2$ drop dramatically and can even become 
statistically acceptable. This implies that these data sets can, 
instead of discarding them, from now on be included in the $np$ database.
The salvation of these systematically-flawed data 
sets~\cite{She74,Hue80,Eri95,Rah98,Fra00} is a major accomplishment
of our adnorm method.

In recent publications~\cite{Ren98,Swa97,Ren97,Ren01} we pointed out that
the Uppsala data at 162 MeV~\cite{Eri95,Rah98} contain unexplained
large systematic errors. 
Also some other $np$ differential cross section measurements
appeared to have systematic errors similar (but not identical) to the
Uppsala data; the Princeton~\cite{She74} and Freiburg~\cite{Hue80,Fra00}
data are prominent examples.

In the following we will compare the data in the standard way with the 
energy-dependent Nijmegen partial-wave analysis PWA93~\cite{Sto93}. 
First of all we will establish that these flawed data sets have significant, 
smoothly angle-dependent, systematic errors.
Since we have no explanation for these systematic experimental errors,
we must simply accept the fact that certain data have such errors.
We are surprised, however, that so many $np$ differential cross sections
have such similar, angle-dependent systematic errors. 
Then we will apply the adnorm method to these data and demonstrate that 
this method can correct for some of such systematic errors.
In order to save two sizable data sets, which became almost acceptable
after using the adnorm method, we changed slightly
the definition of an individual outlier.
This has nothing to do with the adnorm method, but it is a measure taken in
the same spirit: Try to be as frugal as possible with data sets, and do
not omit them from the data base, unless absolutely necessary. 
In this way we obtain normalized data sets that are statistically
acceptable and that can be included henceforth in the $np$ databases
for PWA's.

\section{Normalization}
An $np$ differential cross section $\sigma(\theta,expt)$, 
consisting of $N_{data}$ data points, 
is called {\em experimentally} normalized,
when for this data set the normalization has actually been measured. 
In that case, it has an experimental norm $N(expt)=1.00$ 
with a corresponding experimental error $\delta N(expt)$. 
This norm and error are included as a datum in the data base. 
For backward $np$ scattering this error is often of the order of 4\%
or larger. In the Nijmegen PWA's we also determine for each data set
the normalization $N(pwa)$ with an error $\delta N(pwa)$.
This is a {\em calculated} normalization 
(for a discussion of these points see Ref.~\cite{Ren01}).
This error $\delta N(pwa)$ is in most cases less than 1\%~\cite{Sto93}.
When these two normalizations $N(expt)$ and $N(pwa)$
differ by more than 3 standard deviations (s.d.), 
we remove the experimental normalization and its error from the database. 
The data set is then ``floated'', which means in practice that 
a very large normalization error is assigned to the data set. 
A data set is also floated when its normalization has not been measured
at all. In the case of floated data the calculated normalization $N(pwa)$ is 
determined solely by the angular distribution.
The number of degrees of freedom $N_{df}$ for a set with $N_{data}$
data points is then $N_{df} = N_{data}-1$. 
For the determination of the calculated normalization $N(pwa)$ 
we have sacrificed one degree of freedom.

For seriously flawed data sets one cannot get a sufficiently low
value of $\chi^2$ by merely adjusting or floating the normalization.
Such data sets are then omitted from the data base for PWA's.
The main point of this paper is the observation that for some of
such unacceptable data sets we can introduce 
an {\em angle-dependent} normalization $N(\theta)$ 
in such a way that we essentially sacrifice {\em two or more} degrees 
of freedom to obtain significant drops in the value of $\chi^2$.
Such a sacrifice is unfortunately necessary in our attempt to save these 
data sets from being discarded otherwise.

In the adnorm method we have to make assumptions when we are trying
to parametrize the angular dependence of the systematic errors. 
We first map the experimental angular interval
$[\theta_{min},\theta_{max}]$ onto the interval $[-1,1]$.
This mapping can be done in many ways. 
We consider the two mappings
\begin{equation}
   x = (\theta - \theta_+)/\theta_- \ ,
   \label{eq:map1}
\end{equation}
with $\theta_\pm = (\theta_{max}\pm\theta_{min})/2$, and
\begin{equation}
   x = (\cos\theta - z_+)/z_- \ ,
\label{eq:map2}
\end{equation}
with $z_\pm = (\cos\theta_{max}\pm\cos\theta_{min})/2$.

For the discrete set of $N_{data}$ data points $x_i$ $(i=1,N_{data})$ 
on the $x$-interval $[-1,1]$ 
we define the inner product $(x,y)=(1/N_{data})\sum\nolimits_{i} x_i\,y_i$. 
This allows us to construct the polynomials 
$ S_n(x) = \sum_{i=0}^n a_i \, x^i $, 
which are orthogonal with respect to this inner product 
and normalized such that $a_n=1$.
Next we expand $N(\theta)$ in these orthogonal polynomials $S_n(x)$
on this discrete set of data points. We write
\begin{equation}
    N(\theta) =
    N_0\,\left[ 1 + f(\theta) \sum_{n=1}^p c_n \,S_n(x) \right] \ .
    \label{eq:norm}
\end{equation}
We allow for the introduction of an extra function $f(\theta)$. 
In practical cases we make the simplest choice $f(\theta)\equiv 1$, 
but {\em e.g.} $f(\theta)=1/\sigma(\theta,pwa)$ could also be a suitable choice.
The expansion in orthogonal polynomials gives exactly the
same $N(\theta)$ as a power series expansion up to the same power $p$.
In the case of a power series expansion the coefficients $N_0$ and $c_n$
vary very much with the value of $p$, this is not the case anymore
for an expansion in orthogonal polynomials.
The normalization $N_0$ and the $p$ adnorm parameters $c_n$ ($n=1,p$)
and their errors are determined by the least-squares method, where the
data are compared with PWA93. 
In an actual PWA such data sets contribute with $N_{df}=N_{data}-(1+p)$
degrees of freedom.
The $\chi^2$ that results when $p$ adnorm parameters are introduced is 
called $\chi^2_p$. 
When the standard angle-independent normalization (with zero adnorm parameters) 
is applied, the $\chi^2$ is called $\chi^2_0$.

How many adnorm parameters $c_n$ do we have to introduce? 
The basic rule is that each introduced parameter should cause 
a {\em significant} drop in $\chi^2$. 
We apply a 3 s.d. criterion: We introduce the parameter $c_n$ only
when $\chi^2_{n-1}-\chi^2_n\geq 9$. The parameter is then significant.
This procedure of introducing additional parameters stops 
when no significant drop in $\chi^2$ can be achieved anymore.
 When we end up with nonzero adnorm parameters, then we have shown that 
 there are significant angle-dependent systematic errors present in the 
 data, and we have explicitly parametrized these systematic errors.
   We have convinced ourselves that the final result of the parametrization
   is essentially independent of our specific assumptions. 
   The renormalized differential cross sections, obtained by different ways
   of parametrization, were statistically practically the same.

\section{Uppsala data}
Let us see how this procedure works out for the Uppsala 
data~\cite{Rah98} at $T_L=162$ MeV. 
At this neutron beam energy the $np$ differential cross section 
was measured in five overlapping angular regions. 
We ordered these sets by increasing neutron scattering angles 
and called the sets 1 to 5, where set 1 contains the data at the most 
forward angles and set 5 at the most backward angles~\cite{Ren01}. 
These data were then compared to PWA93. 
We removed the point at 93$^\circ$ from set 2 
because it contributes more than 9 (3 s.d.) to $\chi^2$.

In Table~\ref{tab:1} we list the number of data $N_{data}$ in each set, 
the value $\chi^2_0$ obtained by just applying the standard 
angle-independent normalization (all adnorm parameters $c_n\equiv 0$),
the values of $\chi^2_p(\theta)$ obtained by applying the $\theta$-adnorm 
method of Eq.~\ref{eq:map1}, 
and the values of $\chi^2_p(z)$ obtained by applying the $z$-adnorm 
method of Eq.~\ref{eq:map2}. 

From this Table~\ref{tab:1} we see that the fits of the sets 1 and 4 
improve significantly (a drop in $\chi^2$ of much more than 9) 
when introducing only one adnorm parameter $c_1$. 
For sets 2 and 3 this is not true, and we will therefore take 
$c_1\equiv 0$ for these two sets. 
For set 5 we need two $\theta$-adnorm parameters, 
while only one $z$-adnorm parameter is necessary.
After the $\theta$-adnorm method is applied with only one adnorm parameter
the slope of the angle-dependent normalization $N(\theta)$ is called $\alpha$.
The value of this slope can be found in the next-to-last row of 
Table~\ref{tab:1}. 
This slope is used to compare systematic errors in different experiments.
It is important to note that the slope for set 1 is positive,
while the slopes for the sets 4 and 5 are negative.
In the last row is given the variation in \% of the normalization 
$N(\theta)$ in the case $p$=1 over the interval 
$[\theta_{min},\theta_{max}]$.

The combined data set has $N_{data}=87$ 
and when normalized in the standard angle-independent way 
(no adnorm parameters) we obtain $\chi^2_0=243$. 
This value is 12 s.d. higher than the expectation
value $\langle\chi^2\rangle=82(13)$. 
The value for $\chi^2$ drops to $\chi^2(\theta)=92$ 
after introducing four $\theta$-adnorm parameters
($c_1$ for each of the sets 1 and 4, and $c_1$ and $c_2$ for set 5).
With the $z$-adnorm method it drops to $\chi^2(z)=94$ after introducing
three $z$-adnorm parameters ($c_1$ for each of the  sets 1, 4, and 5).
The difference between the two adnorm methods is minor. 
In order to demonstrate this, 
we calculated the $\chi^2$(dif) for the difference between the two 
differential cross sections obtained by the two adnorm methods. 
This $\chi^2$(dif) is very low, only 0.8 for the total data set. 
The conclusion is that the large systematic errors of unknown origin
present in the Uppsala data can be corrected for by using one of 
the adnorm methods. 
The drop of about 150 in $\chi^2$ resulting from the introduction
of only 3 $z$-adnorm (or 4 $\theta$-adnorm) parameters is impressive.

\begin{figure}[t]
\begin{center}
\includegraphics{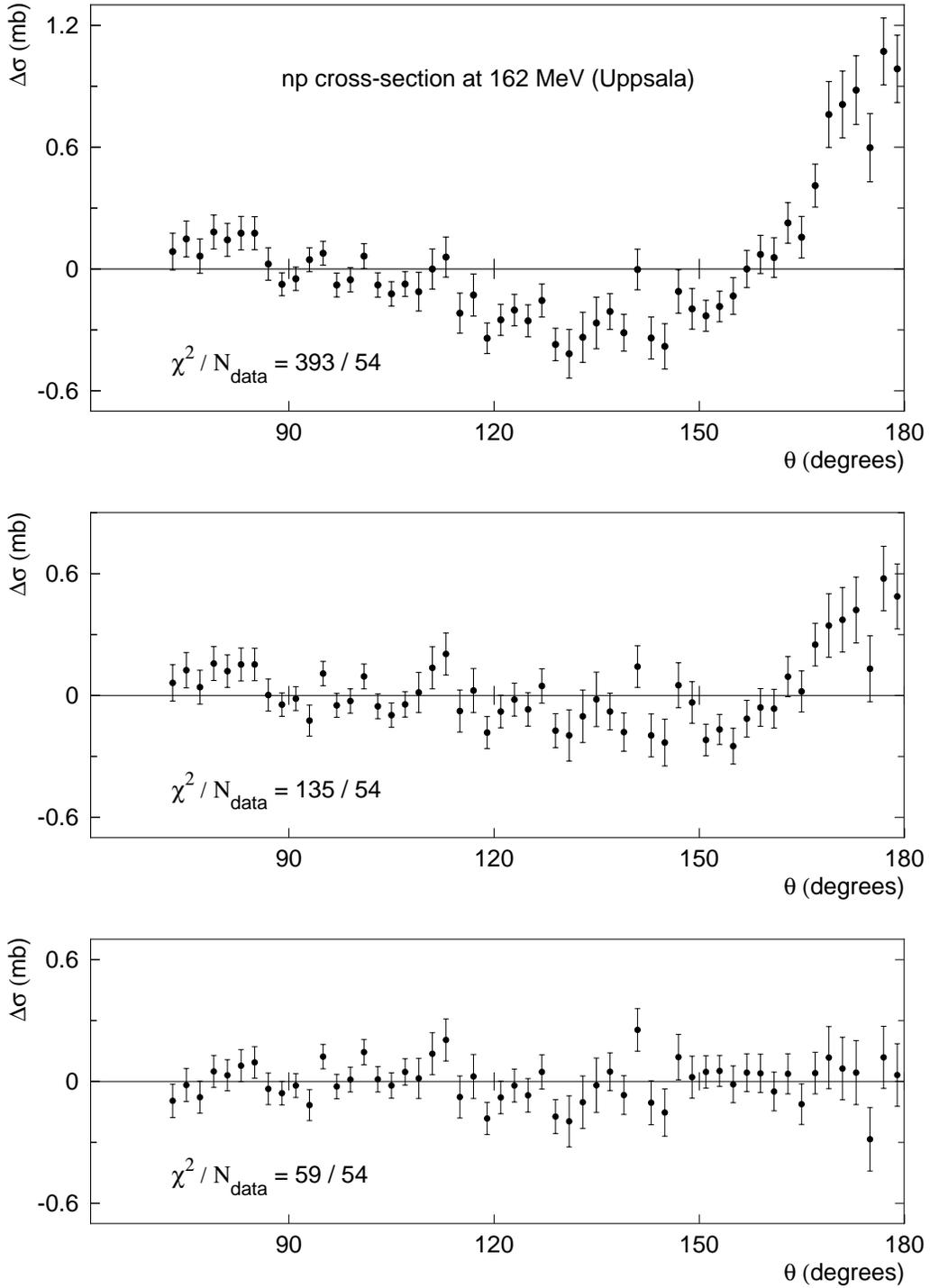}
\end{center}
\caption{Uppsala data at 162 MeV. 
         Top panel: Uppsala's normalization. 
         Middle panel: standard normalization using PWA93.
         Bottom panel: normalized using the adnorm method.}
\label{fig:1}
\end{figure}

To present the data in a similar way as was done by 
the Uppsala group~\cite{Rah98}, 
we averaged the data in the overlap regions between the different sets.
The difference
   $\Delta\sigma(\theta)=N(\theta)\;\sigma(\theta,expt)-\sigma(\theta,pwa)$
normalized in the various ways discussed, is presented in Fig.~\ref{fig:1}.
In the top panel we show the data normalized in the Uppsala way 
and we get $\chi^2=393$ for the 54 data points. 
In the middle panel of Fig.~\ref{fig:1} we show the data normalized 
in the standard angle-independent way. This leads to $\chi^2_0=135$. 
In the bottom panel of Fig.~\ref{fig:1} we present the data normalized 
with the $\theta$-adnorm method. We obtain then $\chi^2=59$. 
From Fig.~\ref{fig:1} one clearly sees the difference between 
the various ways the data have been normalized, and 
the enormous improvement obtained with the adnorm method.

\section{Freiburg data}
Another place where the angle-dependent normalization procedure works
impressively is the abundant Freiburg data~\cite{Hue80,Fra00}. 
This data set consists of 4 different measurements 
(labeled expt I to expt IV) of $np$ differential cross sections, 
each at 20 beam energies between $T_L=199.9$ and 580.0 MeV, 
with a spacing of about 20 MeV. 
Because we compare with PWA93 we can only study data with energies less
than 350 MeV, {\em i.e.} the 8 energies from 199.9 MeV to 340.0 MeV. 
Because of their too high individual contribution to $\chi^2$ 
(more than 3 s.d.) we remove from the database the four data points 
(expt, $T_L$, $\theta$) = (II, 261.9 MeV, $154.96^\circ$),
(II, 300.2 MeV, $148.34^\circ$), (II, 340.0 MeV, $148.07^\circ$), and
(III, 199.9 MeV, $144.32^\circ$).
For the total Freiburg data set we are left then with 859 data points.

In Table~\ref{tab:2} we present the $\chi^2_0/N_{data}$ values for these
4 experiments at 8 energies after the standard angle-independent
normalization. 
It is clear that $\chi^2_0$ for most of these 32 data sets is much too
high. For the total data set of 859 points we find $\chi^2_0=2139$, 
which is 32 s.d. higher than the expectation value 
$ \langle \chi^2_0 \rangle = 827(41) $. 
Therefore, the total Freiburg data set would normally be discarded 
in PWA's. However, we can try the adnorm method. 
The results of applying the $\theta$-adnorm method are presented 
in Table~\ref{tab:3}. 
The first striking observation is the enormous drop in $\chi^2$, 
from 2139 to 831, for the 859 data points. 
This drop was achieved by introducing next to the original 32
normalizations $N_0$ also 45 $\theta$-adnorm parameters. 
This implies on the average a drop of no less than 29 per adnorm parameter.

Looking at the four experiments separately, one sees that 
expt I and expt IV have values for $\chi^2(\theta)$ that 
are smaller than their expectation value ($-0.8$ s.d. and $-0.8$ s.d.), 
that expt II has a $\chi^2(\theta)$ that is 0.6 s.d. higher than its 
expectation value, and that expt III has a $\chi^2(\theta)$ that 
is 3.3 s.d. higher than its expectation value.
This, unfortunately, means that expt III would have to be excluded from
the database for PWA93. 
For the remaining 647 points of expt I, II, and IV we expect 
$\langle\chi^2\rangle=588(34)$. 
We get $\chi^2(\theta)=571$, which is an excellent result. 
We have checked explicitly that the $\chi^2$-distribution of the
renormalized Freiburg data is in very good agreement with the
theoretical expectation~\cite{Ber88}.
With the $z$-adnorm method we got similar results.
Using 46 $z$-adnorm parameters and 32 normalizations we reached
$\chi^2(z)$= 832 for the 859 data points. 
Also in this case expt III should be omitted from the database for PWA's. 
In Table~\ref{tab:3} we also present the values of the 
slope $\alpha$ of the data at $T_L=199.9$ MeV after applying 
the $\theta$-adnorm method with $p=1$.

The conclusion is that the Freiburg data set, 
when compared to PWA93 using the adnorm method, 
has three statistically acceptable experiments and 
one, expt III, that is statistically not acceptable. 
However, one must realize that if this expt III were included in a new PWA, 
then it might possibly have a statistically acceptable value of $\chi^2$.

We would like to point out that also this expt III can be saved, 
when we are willing to bend a little our rule for individual outliers.
In the Introduction we already pointed out that this has nothing to do
with the adnorm method, but only with our wish not to discard data 
unless it is absolutely necessary.
When for expt III a 2.5 s.d. rule is used instead of a 3 s.d. rule, we
must remove also the three data points (III, 199.9 MeV, $133.95^\circ$), 
(III, 240.2 MeV, $149.25^\circ$), and (III, 340.0 MeV, $130.47^\circ$)
as more than 2.5 s.d. outliers. 
In Table~\ref{tab:3} the most right column of expt III contains
the relevant information for this case.
For the 209 data points left from expt III we have the expectation value
$\langle\chi^2\rangle=191(20)$, and we find $\chi^2(\theta)=238$, 
which is 2.3 s.d. higher than expected.
Therefore, expt III is now also statistically acceptable.
The conclusion is that the four Freiburg experiments, consisting of 856 
datapoints, can be made statistically acceptable with the adnorm method
and using the 2.5 s.d. rule for outliers.

\section{Princeton data}
Finally, we consider the Princeton data~\cite{She74}. 
This relatively old data set is generally not included in PWA's and 
these data were {\it e.g.} also discarded in the final version of PWA93. 
We can, however, revisit these data with the adnorm method. 
The results are given in Table~\ref{tab:4}. 
After we have removed two data points 
($T_L=313$ MeV, $\theta=168.1^\circ$ and $170.3^\circ$) as 
more than 3 s.d. outliers, the total set contains 156 data points, 
divided over 9 energies below $T_L=350$ MeV. 
When we normalize these data in the standard manner we get $\chi^2_0=582$. 
This is about 25 s.d. higher than the expectation value. 
Next we applied the $\theta$-adnorm method.
This required 14 additional adnorm parameters. 
The expectation value is then $\langle\chi^2\rangle=133(16)$. 
We obtain $\chi^2(\theta)=195$, which is still 3.9 s.d. too high. 
Therefore the Princeton data, unfortunately, cannot be saved 
by the adnorm method alone, 
despite the enormous improvement in $\chi^2$ from 582 to 195,
which amounts on the average to a drop of 28 per adnorm parameter.
Using the $z$-adnorm method gives similar results.
 
However, when again we are willing to bend our rule for individual
outliers a little, we can also save these data.
According to the 2.5 s.d. rule the three data points ($T_L$, $\theta$) =
(224 MeV, $131.6^\circ$),
(239 MeV, $139.7^\circ$), and
(257 MeV, $178.6^\circ$) must also be omitted.
There are then 153 data points left, which leads to the expectation
value $\langle\chi^2\rangle=130(16)$. 
The second entries in Table~\ref{tab:4} give the relevant information
for this case.
We obtain $\chi^2(\theta)=169$, which is 2.4 s.d. higher than expected.

\section{Discussion and Conclusions}
About the adnorm method that we proposed here, the question could be raised:
``Is it successful because it corrects for experimental errors, or 
perhaps because it corrects for unknown biases in the PWA's?" 
We claim that we correct for unknown systematic {\em experimental} errors.
To demonstrate this we defined the slope parameter 
$\alpha= (1/N_0) (dN(\theta)/d\theta)$ for the case $p=1$.
Looking at the different problematic experiments in about the same
angular region and at about the same energy, we note significant differences.
In the backward direction $[150^\circ,180^\circ]$ there are several
experiments at about the same energy. 
These are the Uppsala set 5 at 162 MeV, the Freiburg expt's I and II 
at 199.9 MeV, and the Princeton data at 182 MeV. The values
of $10^3\:\alpha$ are $-4.9(5)$, $-2.8(4)$, $-2.5(4)$, and $-0.7(4)$,
respectively. These slopes do not agree!
The disagreement between the Uppsala sets 1 and 2 and 
the Freiburg expt IV is worse.
Uppsala set 1 covers the angular region $[73^\circ,107^\circ]$ and 
has $\alpha=3.7(8)\:10^{-3}$.
Uppsala set 2 covers $[89^\circ,129^\circ]$ and has $\alpha=0.3(7)\:10^{-3}$.
The Freiburg expt IV covers $[81^\circ , 124^\circ]$ and has
$\alpha=-1.9(3)\:10^{-3}$, which is even of the {\em opposite} sign as
the values for the Uppsala sets 1 and 2.
When one wants to blame PWA93 for the discrepancy and claim
that the Uppsala and Freiburg data are in agreement, then one
should at least find the same values for $\alpha$. 
Because the $\alpha$ values for these experiments are significantly different,
we can conclude that the Uppsala and Freiburg data are not in agreement with 
each other.
It is then also clear that the discrepancies must be of experimental origin.

In conclusion, we have shown that many of the measurements of the
$np$ differential cross section suffer from similar systematic errors.
These errors are mostly so large that the corresponding data sets cannot be 
included in PWA's.
However, it turned out that these systematic errors have a smooth angular 
dependence, which can easily be parametrized. 
This allowed us then to use the adnorm method to correct for these 
systematic errors. 
In many cases this gave rise to impressive drops in the values 
of $\chi^2$ for several of the seriously flawed data sets. 
Many of the data sets became statistically acceptable after application
of the adnorm method and therefore can now be included in the data base 
for PWA's.
However, some of the sets required also a slight change in the definition 
of outlier to make them acceptable.
In this manner, several $np$ data sets~\cite{Eri95,Rah98,Fra00,She74}
can be saved from oblivion.

\acknowledgments
We thank M.C.M. Rentmeester, Th.A. Rijken, and R.A. Bryan for helpful 
discussions.
The research of R.G.E.T. was made possible by a fellowship of 
the Royal Netherlands Academy of Arts and Sciences (KNAW).


\newpage
\begin{table}
\begin{tabular}{l|ccccc|r}
    set         &  1    &   2   &    3    &    4    &    5    & total \\ \hline
 $N_{data}$     & 18    &  20   &   18    &   16    &   15    &   87  \\
 $\chi^2_0$     & 38    &  35   &   18    &   35    &  117    &  243  \\
 $\chi^2(\theta)/p$ & 15/1  &  35/0 &  18/0   &  15/1   &   9/2   &   92  \\
 $\chi^2(z)/p$      & 15/1  &  35/0 &  18/0   &  16/1   &  10/1   &   94  \\
 $10^3\,\alpha$ & 3.7(8)& 0.3(7)&$-1.5(6)$&$-2.2(5)$&$-4.9(5)$&       \\
  var in \%     & 12.8  &  3.2  &   5.4   &  7.0    &  12.2   &
\end{tabular}
\caption{Results for the Uppsala data at 162 MeV.}
\label{tab:1}
\end{table}

\begin{table}
\begin{tabular}{c|cccc}
  $T_L$ (MeV)     &  expt I &  expt II& expt III& expt IV  \\ \hline
   199.9          &  71/27  &  58/27  &  44/25  &  57/22   \\
   219.8          &  66/27  &  42/28  &  71/27  &  64/22   \\
   240.2          &  76/27  &  50/30  &  53/27  &  73/23   \\
   261.9          &  82/27  &  40/30  &  65/27  & 122/23   \\
   280.0          &  68/27  &  56/32  &  57/26  & 132/24   \\
   300.2          &  65/27  & 102/32  &  74/27  &  62/24   \\
   320.1          &  47/27  &  63/33  &  82/26  &  70/24   \\
   340.0          &  47/27  &  60/33  &  70/27  &  50/24   \\ \hline
   total          & 522/216 & 471/245 & 516/212 & 630/186  \\
\end{tabular}
\caption{The values of $\chi^2_0/N_{data}$ for the 32 
         Freiburg data sets [$T_L$,expt] and the totals per experiment.}
\label{tab:2}
\end{table}

\begin{table}
\begin{tabular}{c|cccc}
 $T_L$ (MeV) & expt I  & expt II & expt III   & expt IV \\ \hline
 199.9       &  29/1   &  21/1   &  44/0/37   &  13/2   \\  
 219.8       &  26/2   &  23/1   &  31/1/31   &  16/2   \\ 
 240.2       &  23/2   &  30/1   &  28/1/21   &  18/2   \\
 261.9       &  24/2   &  16/2   &  27/2/27   &  32/1   \\
 280.0       &  15/2   &  36/1   &  26/1/26   &  19/1   \\   
 300.2       &  24/2   &  44/1   &  31/2/31   &  25/1   \\
 320.1       &  20/1   &  36/2   &  33/2/33   &  16/1   \\
 340.0       &  18/2   &  33/1   &  40/1/32   &  14/1   \\ \hline
 total       & 179/14  & 239/10  & 260/10/238 & 153/11  \\
$\langle\chi^2(\theta)\rangle$ & 194(20) & 227(21) & 194(20)191 & 167(18)  \\
  s.d.       & $-0.8$  &  0.6    & 3.3\;/\;2.3 & $-0.8$  \\
$10^3\,\alpha$ &$-2.8(4)$&$-2.5(4)$&$-1.1(5)$&$-1.9(3)$ \\
\end{tabular}
\caption{The values of $\chi^2(\theta)/p$ ($p$ is the number of adnorm 
         parameters) for the 32 Freiburg data sets [$T_L$,expt], 
         the totals per experiment, their expectation value, 
         and the slope $\alpha$.
         The last column for expt III gives $\chi^2(\theta)$ after three
         additional data points were removed.}
\label{tab:3}
\end{table}

\begin{table}
\begin{tabular}{c|ccccr}
$T_L$(MeV)&$N_{data}$&$\chi^2_0$& $p$ &$\chi^2(\theta)$&$10^3\,\alpha$\\ \hline
  182      &   14     &    11    &  0  &   11     & $-0.7(4) $  \\  
  196      &   16     &    49    &  2  &   27     & $-1.2(3) $  \\ 
  210      &   16     &    43    &  1  &   14     & $-1.9(4) $  \\
  224      &  16/15   &   71/67  &  1  &  24/16   & $-2.6(4) $  \\
  239      &  18/17   &   46/37  &  1  &  22/14   & $-1.9(4) $  \\   
  257      &  19/18   &   90/61  &  2  &  32/22   & $-1.7(3) $  \\
  284      &   19     &   114    &  2  &   21     & $-2.4(3) $  \\
  313      &   17     &    76    &  2  &   20     & $-2.2(3) $  \\ 
  344      &   21     &    82    &  3  &   24     & $-2.1(4) $  \\ \hline 
  total    & 156/153  &  582/540 & 14  & 195/169  &             \\
\end{tabular}
\caption{Results for the Princeton data.}
\label{tab:4}
\end{table}

\end{document}